\documentclass[prd,twocolumn,floats,floatfix,nofootinbib]{revtex4}
\usepackage[dvips]{graphicx}
\usepackage{graphics}
\usepackage{dcolumn}
\usepackage{amssymb}
\usepackage{bm}
\usepackage{amsmath}

\def\spose#1{\hbox to 0pt{#1\hss}}

\def\lta{\mathrel{\spose{\lower 3pt\hbox{$\mathchar"218$}}
     \raise 2.0pt\hbox{$\mathchar"13C$}}}
\def\gta{\mathrel{\spose{\lower 3pt\hbox{$\mathchar"218$}}
     \raise 2.0pt\hbox{$\mathchar"13E$}}}
\newcommand{\be}{\begin{equation}}
\newcommand{\en}{\end{equation}}
\newcommand{\bea}{\begin{eqnarray}}
\newcommand{\ena}{\end{eqnarray}}

\newcommand{\dd}{\mbox{d}}

\newcommand{\ie}{\textsl{i.e.~}}

\begin{document}
\title{Large classical universes emerging from quantum cosmology}

\author{Nelson Pinto-Neto}
\affiliation{ICRA - Centro Brasileiro de
Pesquisas F\'{\i}sicas -- CBPF, \\ rua Xavier Sigaud, 150, Urca,
CEP22290-180, Rio de Janeiro, Brazil}

\date{\today}

\begin{abstract}

It is generally believed that one cannot obtain a large Universe from
quantum cosmological models without an inflationary
phase in the classical expanding era because the typical size
of the Universe after leaving the quantum regime should be around
the Planck length, and the standard decelerated classical expansion after that is not
sufficient to enlarge the Universe in the time available. For instance,
in many quantum minisuperspace bouncing models studied in the literature,
solutions where the Universe leave the quantum 
regime in the expanding phase with appropriate size have negligible probability amplitude
with respect to solutions leaving this regime around the
Planck length. In this paper,
I present a general class of moving gaussian solutions of
the Wheeler-DeWitt equation where the velocity of the wave
in minisuperspace along the scale factor axis, which is the new large
parameter introduced in order to circumvent the abovementioned problem, 
induces a large acceleration around the quantum bounce,
forcing the Universe to leave the quantum regime sufficiently
big to increase afterwards to the present size, without needing any classical inflationary
phase in between, and with reasonable relative probability amplitudes with respect
to models leaving the quantum regime around the Planck scale. Furthermore,
linear perturbations around this background model are free of any transplanckian problem.


\end{abstract}

\maketitle

\section{Introduction}

The existence of an initial singularity~\cite{singularity} is one of the major drawbacks of
classical cosmology. In spite of the fact that the standard cosmological
model, based in classical general relativity sourced by
ordinary matter, has been
successfully tested until the nucleosynthesis era,
the extrapolation of this model to higher energies leads to a breakdown
of the geometry in a finite cosmic time. It
indicates the failure of this conventional approach at high energies,
which should be complemented through the intervention of some new physics
(presence of exotic matter, modifications of general relativity through 
non-minimal couplings, non linear curvature terms in the lagrangian,
quantum effects of the gravitational field , etc), leading to a complete 
regular cosmological model.

In the framework of quantum cosmology in minisuperspace models,
non singular bouncing models have been obtained\footnote{There are many other 
frameworks where bounces connecting the present 
expanding phase with a preceding contracting one may occur~\cite{bounce0,bounce}. 
In this case, the
Universe is eternal, there is no beginning of time, nor horizons.
The new features of these models
introduce a new picture, where the usual problems of initial conditions \cite{noinf}
(as, for instance, the almost homogeneous beginning of the expanding phase might
be explained through the disipation of nonlinear inhomogeneities when the universe
was very large and rarefied in the asymptotic far past of the contracting phase)
and the evolution of cosmological perturbations \cite{pertbounce}
are viewed from a very different perspective.}, where the bounce occurs due
to quantum effects in the
background~\cite{qc,bohmqc,fabris}. Some
approaches have used an ontological interpretation of quantum
mechanics, the Bohm-de Broglie~\cite{dbb} one, to interpret the
results~\cite{bohmqc,fabris} because, contrary to the standard Copenhagen
interpretation, this ontological interpretation does not
need a classical domain outside the quantized system to generate the
physical facts out of potentialities (the facts are there {\it ab
initio}), and hence it can be applied to the Universe as a
whole. Of course there are other alternative interpretations which can be used in quantum
cosmology, like the many worlds interpretation of quantum mechanics~\cite{everett},
but I will not use them in this paper.

In the Bohm-de Broglie interpretation,
quantum Bohmian trajectories, the quantum evolution of the scale
factor $a_q(t)$, can be defined through the
relation $\dot{a}\propto\partial S/\partial a$, where $S$ is 
the phase of an exact wave solution $\Psi(a,t)$ of the
Wheeler-DeWitt equation. It satisfies a modified
Hamilton-Jacobi equation, augmented with a quantum potential term
derived from $\Psi(a,t)$, and hence $a_q(t)$ is not the classical
trajectory: in the regions where the quantum effects cannot be
neglected, the quantum trajectory $a_q(t)$ performs a bounce which
connect two asymptotic classical regions where the quantum effects
are negligible. One then has in hands a definite function of time
for the homogeneous and isotropic background part of the Universe, 
even at the quantum level, which realizes a
soft transition from the contracting phase to the expanding one.

When studying the evolution of quantum cosmological perturbations
on these backgrounds, which was done in the series of papers 
\cite{PPN1,PPN2,PPN3,PPN4} for the case of one perfect fluid
with equation of state $p=w\rho$, one arrives at the result that, in order
to obtain wavelength spectra and amplitudes compatible with
CMB data, one must have \cite{PPN4} $|w| << 1$ and $w^{1/4}L_0\approx 10^2 l_{\rm pl}$,
where $L_0$ is the curvature scale at the bounce and $l_{\rm pl}$ is the Planck
length. Hence this analysis shows that the model is self consistent because observational constraints
impose that the curvature scale at the bounce must be at least a few orders
of magnitude greater than the Planck length, a region where one can trust
the Wheeler-DeWitt equation without been spoiled by high order quantum gravity
effects. Of course this model should be extended to include radiation. 
In Ref.~\cite{noinf} it is shown that the requirement $|w| << 1$ is important
only at the moment when the perturbation wavelength becomes greater than the
curvature scale: the fluid which dominates at the bounce is irrelevant for the
spectral index (but is important for the amplitude, as we will see in future publications). 

However this scenario has a problem on the quantum background solution itself:
in order for the model describe the big Universe we live in, the scale
factor at the bounce $a_0$ must be somewhat large, and the probability one
can obtain from the wave function of the model for the occurence of
this value is incredibly small (in some cases $\exp(-10^{89})$, as we
will see later on). Hence, either there is an inflationary phase after the
bounce in order to enlarge the Universe from a small $a_0$
(which may lead to transplanckian problems~\cite{transP} and non linear inhomogeneities
at the bounce because of the growth of linear perturbations in the contracting
phase if $a_0$ is small), or one should rely very strongly on some anthropic principle
in a situation much worst than in the landscape scenario.

The aim of this paper is to overcome this difficulty by proposing
some more general wave solutions of the Wheeler-DeWitt equation which
lead to realistic bouncing scenarios with parameters with reasonable 
probabilities and without any transplanckian problem. In Ref. \cite{PPN4},
the wave function at the bounce was chosen to be a static gaussian of the 
scale factor centered at $a=0$. As we will see, the ratio between the scale
factor at the bounce $a_0$ and the width of the gaussian must be very large
in order to yield the big Universe we live in, yielding the very small probability
of occurrence of these parameters I mentioned above. However, if one generalizes the wave function
to be a moving gaussian on the $a$-axis with velocity $u$, there is a minimum value of
this parameter from where one can obtain a large Universe, with reasonable probability 
of occurrence, and without any transplanckian problems.
The parameter $u$ induces a very large acceleration around the bounce, leading
to a sufficiently large scale factor when the quantum regime is over.

This paper is organized as follows: in Section II I describe in detail the problem I want
to solve. In Section III I present the generalized wave solutions
from which this problem can be circumvented. I conclude in Section IV with a discussion
of our results, their physical meanings, and prospects for future work.

\section{Quantum bounce solutions from static initial gaussians and their problems}

The Hamiltonian constraint describing a cosmological model with flat homogeneous and isotropic
closed spacelike hypersurfaces with comoving volume $V=1$,
and a perfect fluid satisfying $p=\omega\epsilon$, where
$\epsilon$ is the perfect fluid energy density, $p$ is the pressure and 
$\omega$ is a constant, reads \cite{PPN1}
\begin{equation}
\label{h00} H_0\equiv
\frac{P_{T}}{a^{3\omega}}-\frac{P_{a}^{2}}{4a},
\end{equation}
where $a$ is the scale factor, $P_a$ its canonical momentum, and the conserved 
quantity $P_T$, the momentum canonically conjugated to the degree of freedom
of the fluid $T$ \cite{rubakov}, is associated with the constant appearing in the energy density
of the fluid through the relation $\epsilon = P_T/a^{3(1+w)}$. 
All quantities are in Planck unities. One can verify
that the Hamiltonian $H=NH_0$ generates the usual Friedmann equations of the model.

The wave function $\Psi(a,T)$ satisfies the Wheeler-DeWitt equation $H_0\Psi=0$,

\begin{equation}
\label{schroedinger-separado-fundo}  i  \frac{\partial}{\partial
T} \Psi(a,T)=\frac{a^{(3\omega-1)/2}}{4}
\frac{\partial}{\partial a} \left[
a^{(3\omega-1)/2}\frac{\partial}{\partial a}\right] \Psi(a,T),
\end{equation}
where I have chosen the factor ordering in $a$ in order to yield a
covariant Schr\"odinger equation under field redefinitions.
The fluid selects a prefered time variable.

I change variables to
$$\chi=\frac{2}{3} (1-\omega)^{-1} a^{3(1-\omega)/2},$$ obtaining the
simple equation
\begin{equation}
i\frac{\partial\Psi(\chi,T)}{\partial T}= \frac{1}{4}
\frac{\partial^2\Psi(\chi,T)}{\partial \chi^2}. \label{es202}
\end{equation}
This is just the time reversed Schr\"odinger equation for a one
dimensional free particle constrained to the positive axis. As $a$
and $\chi$ are positive, solutions which have unitary evolution must
satisfy the condition
\begin{equation}
\label{cond27} \biggl(\Psi^{\star}\frac{\partial\Psi}{\partial
\chi} -\Psi\frac{\partial\Psi^{\star}}{\partial
  \chi}\biggr)\Biggl|_{\chi=0}=0
\end{equation}
(see Ref.~\cite{fabris} for details). I can choose the initial
normalized wave function
\begin{equation}
\label{initial}
\Psi^{(\mathrm{init})}(\chi)=\biggl(\frac{8}{T_0\pi}\biggr)^{1/4}
\exp\left(-\frac{\chi^2}{T_0}\right) ,
\end{equation}
where $T_0$ is an arbitrary constant. The gaussian
$\Psi^{(\mathrm{init})}$ satisfies condition (\ref{cond27}), and it 
gives the probability density for the value of $\chi$ at $T=0$
with minimum uncertainty.

Using the propagator procedure of Refs.~\cite{fabris}, we
obtain the wave solution for all times in terms of $a$:
\begin{widetext}
\begin{equation}\label{psi1t}
\Psi(a,T)=\left[\frac{8 T_0}{\pi\left(T^2+T_0^2\right)}
\right]^{1/4}
\exp\biggl[\frac{-4T_0a^{3(1-\omega)}}{9(T^2+T_0^2)(1-\omega)^2}\biggr]
\exp\left\{-i\left[\frac{4Ta^{3(1-\omega)}}{9(T^2+T_0^2)(1-\omega)^2}
+\frac{1}{2}\arctan\biggl(\frac{T_0}{T}\biggr)-\frac{\pi}{4}\right]\right\}
.
\end{equation}
\end{widetext}

Due to the chosen factor ordering, the probability density
$\rho(a,T)$ has a non trivial measure and it is given by
$\rho(a,T)=a^{(1-3\omega)/2}R^2$, where $R^2=\left|\Psi(a,T)\right|^2$.  Its
continuity equation, one of the equations coming from Eq.~(\ref{schroedinger-separado-fundo}) 
after substitution of $\Psi=Re^{iS}$ in it, reads
\begin{equation}
\label{cont} \frac{\partial\rho}{\partial T}
-\frac{\partial}{\partial a}\biggl[\frac{a^{(3\omega-1)}}{2}
\frac{\partial S}{\partial a}\rho\biggr]=0 ,
\end{equation}
which implies, in the Bohm interpretation \cite{dbb}, the definition of a velocity field
\begin{equation}
\label{guidance} \dot{a}=-\frac{a^{(3\omega-1)}}{2} \frac{\partial
S}{\partial a} ,
\end{equation}
in accordance with the classical relations $\dot{a}=\{a,H\}=
-a^{(3\omega-1)}P_a/2$ and $P_a=\partial S/\partial a$.

Note that $S$ satisfies
the other equation coming from (\ref{schroedinger-separado-fundo}),
\begin{equation}
\label{hamilton-jacobi}  
\frac{\partial S}{\partial T}-\frac{a^{(3\omega-1)}}{4}
\biggl(\frac{\partial S}{\partial a}\biggr)^2 +
\frac{a^{(3\omega-1)/2}}{4R}
\frac{\partial}{\partial a} \left[
a^{(3\omega-1)/2}\frac{\partial R}{\partial a}\right]=0 ,
\end{equation}
which is a Hamilton-Jacobi-like equation with an extra quantum term,
called the quantum potential, given by
\begin{equation}
\label{quantumpotential}  
Q\equiv -\frac{a^{(3\omega-1)/2}}{4R}
\frac{\partial}{\partial a} \left[
a^{(3\omega-1)/2}\frac{\partial R}{\partial a}\right] .
\end{equation}
Hence, the trajectory (\ref{guidance}) will not coincide with the classical trajectory
whenever $Q$ is comparable with the other terms present in Eq.~(\ref{hamilton-jacobi})
because $S$ will be different from the classical Hamilton-Jacobi function.

Inserting the phase of (\ref{psi1t}) into Eq.~(\ref{guidance}), I
obtain the Bohmian quantum trajectory for the scale factor:
\begin{equation}
\label{at} a(T) = a_0
\left[1+\left(\frac{T}{T_0}\right)^2\right]^\frac{1}{3(1-\omega)} ,
\end{equation}
or, in terms of $\chi(T)$,
\begin{equation}
\label{xt} \chi(T) = \chi_0
\left[1+\left(\frac{T}{T_0}\right)^2\right]^\frac{1}{2} .
\end{equation}
Note that $\chi_0$ is the value of $\chi$ at $T=0$, the moment of the bounce, $\chi_0=\chi_{\rm bounce}$, 
and the scale factor at the bounce $a_0$ is connected to $\chi_0$
through
\begin{equation}
\label{a0}
a_0=\biggr[\frac{3}{2} (1-\omega) \chi_0\biggl]^{2/[3(1-\omega)]}.
\end{equation}

Solution (\ref{at}) has no singularities and tends to the
classical solution when $T\rightarrow\pm\infty$. Remember that I
am in the gauge $N=a^{3\omega}$, and $T$ is related to conformal
time through
\begin{equation}
\label{jauge} N\dd T = a \dd \eta \quad \Longrightarrow \dd\eta =
\left[a(T)\right]^{3\omega-1} \dd T.
\end{equation}
The solution (\ref{at}) can be obtained from other initial wave
functions (see Ref.~\cite{fabris}).

However, the above solution suffers from the following drawback:
the curvature scale at the bounce reads $L_{\rm bounce}\equiv T_0a_0^{3w}$, and
the quantity $P_T$ associated in the classical limit $|T|\rightarrow\infty$ with the constant 
appearing in the energy density
of the fluid through the relation $\epsilon = P_T/a^{3(1+w)}$, can be
obtained in the Bohmian approach
from the wave function through the relation $P_T=\partial S/\partial T$.
It reads
\begin{equation}
P_T=\frac{\partial S}{\partial T}=\frac{T_0}{2(T^2+T_0^2)}-
\frac{\chi(T)^2(T_0^2-T^2)}{(T^2+T_0^2)^2}.
\label{pt1}
\end{equation}
Inserting the solution (\ref{xt}) in Eq.~(\ref{pt1}) and taking 
the classical limit $|T|\rightarrow\infty$, one obtains
\begin{equation}
P_T=\frac{\chi_0^2}{T_0^2}.
\label{pt100}
\end{equation}

In the case of dust,
$P_T$ is the total dust mass of the Universe, yielding $P_T\geq 10^{60}$.
If one takes the curvature scale at the bounce some few orders of magnitude
larger than the Planck length, say $10^3$, in order to not spoil the Wheeler-DeWitt approach
used above due to strong quantum gravitational effects, one has to have
$\chi_0\geq 10^{33}$, with probability less than $\exp(-10^{63})$ to occur (see Eq.~(\ref{initial})).

The situation is similar with radiation, where now $P_T = \chi_0^2/T_0^2 \geq 10^{116}$.
Note that in this case $\chi_0=a_0$, $T=\eta$, and the curvature scale at the bounce reads 
$L_{\rm bounce}\equiv T_0a_0$.
Combining the constraints $a_0/T_0 \geq 10^{58}$ and $a_0T_0\geq 10^3$, one arrives
at the very low probability $\exp(-10^{89})$ for these parameters to occur.

The source of the problem is the fact that the constant $\chi_0$ appearing in Eq.~(\ref{pt100})
is also the value of $\chi$ at $T=0$, the $\chi$ at the bounce, and the fact that $P_T$ 
must be large, induces
a large $\chi^2/T_0$ in the gaussian (\ref{initial}). One possibility to scape from
this drawback is to find a different
wave solution to Eq.~(\ref{schroedinger-separado-fundo}) which either modifies Eq.~(\ref{pt100}) 
or yields Bohmian trajectories where $\chi_0$ is not anymore
the value of $\chi$ at $T=0$, allowing the possibility of having a small initial $\chi$,
hence a small $\chi^2/T_0$ in (\ref{initial}), and a huge $\chi_0$, perhaps through
the presence of an inflationary phase between the bounce and the standard decelerated expansion. 
I will show in the next section that it is indeed possible to obtain a more general class of wave solutions
of Eq.~(\ref{schroedinger-separado-fundo}) where the above mentioned problem
is circumvented.

\section{New bouncing solutions}

I will generalize the initial wave function by inserting 
a velocity term in Eq.~(\ref{initial}), which of course must satisfy the boundary condition
(\ref{cond27}), and now reds,
\begin{widetext}
\begin{equation}
\label{initial2}
\Psi^{(\mathrm{init})}(\chi)=\biggl(\frac{2}{T_0\pi}\biggr)^{1/4}[1\pm\exp(-u^2T_0/8)]^{-1/2}
[\exp(iu\chi/2)\pm\exp(-iu\chi/2)]\exp\left(-\frac{\chi^2}{T_0}\right) .
\end{equation}
This initial wave function represents two gaussians travelling from the origin in opposite
directions (keeping in mind that only the tail of the 
gaussian traveling in the negative direction with suport on the positive $a$ axis has
physical meaning). 
The solution for all times read

\begin{eqnarray}\label{psi1tu}
\Psi(\chi,T)&=&\left[\frac{2 T_0}{\pi\left(T^2+T_0^2\right)}
\right]^{1/4}(1\pm\exp(-u^2T_0/8)^{-1/2}\nonumber\\
&\biggl(&\exp\biggl[-\frac{T_0(\chi-uT)^2}{(T^2+T_0^2)}\biggr]
\exp\left\{-i\left[\frac{T(\chi-uT)^2}{(T^2+T_0^2)}+2u(\chi-uT/2)
+\frac{1}{2}\arctan\biggl(\frac{T_0}{T}\biggr)-\frac{\pi}{4}\right]\right\}\nonumber\\
&\pm&\exp\biggl[-\frac{T_0(\chi+uT)^2}{(T^2+T_0^2)}\biggr]
\exp\left\{-i\left[\frac{T(\chi+uT)^2}{(T^2+T_0^2)}-2u(\chi+uT/2)
+\frac{1}{2}\arctan\biggl(\frac{T_0}{T}\biggr)-\frac{\pi}{4}\right]\right\}\biggr) ,
\end{eqnarray}
which I write as, 
\[
\Psi=A(R_{-}e^{i S_{-}}\pm R_{+}e^{i S_{+}}) ,
\]
where
\begin{eqnarray*}
R_{\pm}&\equiv &\exp\biggl[-\frac{T_0(\chi \pm uT)^2}{(T^2+T_0^2)}\biggr],\nonumber \\
S_{\pm}&\equiv &\left[-\frac{T(\chi\pm uT)^2}{(T^2+T_0^2)}\pm 2u(\chi\pm uT/2)
-\frac{1}{2}\arctan\biggl(\frac{T_0}{T}\biggr)+\frac{\pi}{4}\right],\nonumber\\
A&\equiv& \left[\frac{2 T_0}{\pi\left(T^2+T_0^2\right)}
\right]^{1/4}(1\pm\exp(-u^2T_0/8)^{-1/2}.
\end{eqnarray*}
From these equations one obtains the total amplitude and phase as
\begin{eqnarray*}
\emph{R}&=A&\sqrt{R_{+}^{2}+R_{-}^{2}\pm2R_{+} R_{-}\cos(S_{+}-S_{-})}\\
\emph{S}&=&{\rm arctan}\left(\frac{ R_{+}\sin(S_{+})\pm R_{-}\sin(S_{-}) }{ R_{+}\cos(S_{+})\pm R_{-}\cos(S_{-})  }\right)
\end{eqnarray*}
The derivative of $S$ with respect to some variable $x$ reads
\begin{eqnarray*}
\label{flu}
\frac{\partial \emph{S} }{\partial x }&=&\frac{
R^{2}_{+}\frac{\partial S_{+}}{\partial x }+R^{2}_{-}\frac{\partial S_{-}}{\partial x }\pm\left(\frac{\partial S_{+}}{\partial x }+\frac{\partial S_{-}}{\partial x } \right)R_{+}R_{-}\cos\left(S_{+}-S_{-}\right)\pm\left(R_{-}\frac{\partial R_{+}}{\partial x }-R_{+}\frac{\partial R_{-}}{\partial x }\right)\sin\left(S_{+}-S_{-}\right)}{R_{+}^{2}+R_{-}^{2}\pm2R_{+} R_{-}\cos(S_{+}-S_{-})}
\end{eqnarray*}

The guidance relation (\ref{guidance}) leads to the exact differential equation 

\be
\label{exact}
\dot{\chi}(T)=\frac{T\chi(T)}{T^2+T_0^2}+\frac{uT_0^2}{T^2+T_0^2}
\biggl[\frac{\sinh\biggl(\theta \frac{T}{T_0}\biggr)\pm\frac{T}{T_0}\sin\theta}{\cosh\biggl(\theta \frac{T}{T_0}\biggr)\pm\cos\theta}\biggr],
\en
where  
\be
\theta\equiv\frac{4uT_0^2\chi(T)}{T^2+T_0^2}.
\en

From the solution (\ref{psi1tu}), the new $P_T$ is now given by 
\begin{equation}
\label{pt2}
P_T=\frac{\partial S}{\partial T}=
\frac{T_0}{2(T^2+T_0^2)}+
\frac{[u^2T_0^2-\chi ^2(T)](T_0^2-T^2)}{(T^2+T_0^2)^2}+\frac{\biggl[4uT_0^2T\chi(T)\sinh\biggl(\theta \frac{T}{T_0}\biggr)\pm
2T_0u\chi(T)(T^2-T_0^2)\sin\theta\biggr]}{(T^2+T_0^2)^2\biggl[\cosh\biggl(\theta \frac{T}{T_0}\biggr)
\pm \cos\theta\biggr]}.
\end{equation}

From now on I will work with the plus sign solution given in Eq. (\ref{psi1tu}). The minus
sign solution yields the same qualitative results.
\end{widetext}

\subsection{Quantum solutions for small $u$}

Taking $u<<1$, one has 
\be
\label{eqinflation}
\dot{\chi}(T)=\frac{T\chi(T)}{T^2+T_0^2}+\frac{4\chi(T)u^2TT_0^3}{(T^2+T_0^2)^2} ,
\en
with solution
\begin{eqnarray}
\label{inflation}
\chi(t)&=&\frac{\chi_0}{T_0}\sqrt{T^2+T_0^2}\exp\biggl[\frac{-2u^2T_0^3}{T^2+T_0^2}\biggr]\nonumber\\
&=&\chi_0\sqrt{x^2+1}\exp\biggl[\frac{-2u^2T_0}{x^2+1}\biggr],
\end{eqnarray}
where in the last step I wrote the solution in terms of $x=T/T_0$.

Solution (\ref{inflation}) has very nice properties. First of all, one can see that
the values of $\chi$ and the curvature scale at the bounce are now given by
\be
\label{chibounce}
\chi_{\rm bounce}=\chi_0\exp(-2u^2T_0),
\en
and
\be
\label{lcbounce}
L_{\rm bounce}=\exp\biggl[-\frac{4wu^2T_0}{(1-w)}\biggl]\frac{T_0a_0^{3w}}{\sqrt{1+4u^2T_0}}.
\en
Inserting solution (\ref{inflation}) into
Eq.~(\ref{pt2}) in the limit $|T|\rightarrow\infty$, yields
\begin{equation}
P_T=\frac{\chi_0^2}{T_0^2}.
\label{pt11}
\end{equation}

Note that now $\chi_0\neq\chi_{\rm bounce}$, the value of $\chi$ at $T=0$. 
In fact, from Eq.~(\ref{chibounce}),
one may have $\chi_0 >>\chi_{\rm bounce}$, depending on the value of $T_0$, because of the
huge acceleration one may obtain near after the bounce as compared with the case where
$u=0$: it is a bounce followed by inflation. Hence, one may have reasonable probability
amplitudes for the free parameters of the theory which are compatible with a huge $P_T$.
 
However, one must also check whether the Bohmian trajectory (\ref{inflation}),
with such apropriate choice of parameters, reachs classical evolution ($x>>1$)
before the nucleosynthesis epoch.
Let us concentrate on the case of radiation ($w=1/3$), as it is the most interesting physical
situation (one expects the quantum effects and the bounce to occur in a very hot radiation dominated
universe). Supose the classical limit is already valid at a conformal time where the energy
density of radiation is minimally greater than the energy density before nucleosynthesis,
say, the energy density around the freeze-out of neutrons, $\rho_{\rm f}\approx 10^{-88}$. 
Then, from Eqs.~(\ref{inflation}) and (\ref{pt11}),
and from $\rho_{\rm f}=P_T/a_{\rm f}^4$, with $a_{\rm f}=a_0\eta_{\rm f}/T_0$, one obtains 
$x_{\rm f}\equiv\eta_{\rm f}/T_0\approx 10^{22}/(T_0P_T^{1/4})$. Using that $P_T\geq 10^{116}$,
one gets that $x_{\rm f}\leq 10^{-7}/T_0$. Hence, $x_{\rm f}>>1$ if and only if
$T_0<<10^{-7}$. However, as $u<<1$, then $u^2T_0<<1$, and the exponential in (\ref{inflation})
would be irrelevant, turning solution (\ref{inflation})
very close to solution (\ref{at}) for $w=1/3$, taking us back to the previous problem.
Concluding, the only way to obtain a huge $P_T$ with parameters with reasonable probability
amplitudes in this framework is through choices which will
change the usual scale factor evolution during nuclosynthesis, spoiling its observed predictions.

\subsection{Quantum solutions for large $u$}

If $u>>1$, for $|T|$ not very small, and noting that the unique possible
asymptotic behaviour of a solution $\chi(T)$ of Eq.~(\ref{exact}) is $\chi(T)\propto T$, then 
the hyperbolic functions in (\ref{exact}) are very large and much
greater than the terms with trigonometric functions, yielding 
\be
\label{equ}
\dot{\chi}(T)=\frac{T\chi(T)}{T^2+T_0^2}\pm\frac{uT_0^2}{T^2+T_0^2},
\en
with solution
\be
\label{u}
\chi(T)=\frac{\chi_0}{T_0}\sqrt{T^2+T_0^2}\pm uT,
\en
where the $\pm$ sign corresponds to positive and negative values of $T$, respectively.
For $|T|\approx 0$, one has to rely on numerical calculations. However, as shown in
figure 1 below, for large $u$ this difference is quite unimportant. Hence, again, $\chi_0$ 
is very close to the value of $\chi$ at the bounce.

\begin{figure}
\includegraphics[angle=-90,width=8cm]{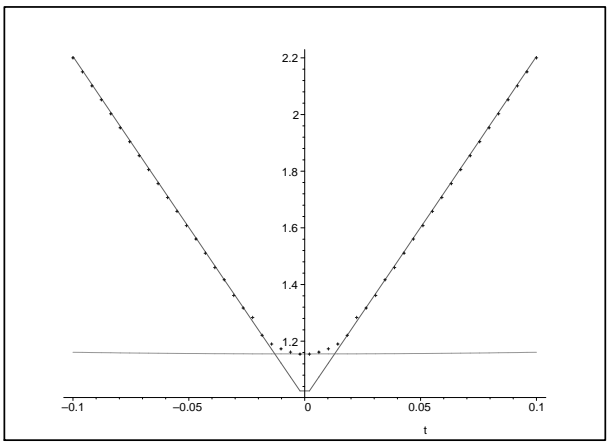}
\caption{Figure showing $\chi(t)$ against $t$, comparing solution (\ref{u}) (continuous line) for
$u=12,\chi_0=1,T_0=1$ with the numerical solution of equation
(\ref{exact}) for $u=12$, $\chi(10)=130$ (dotted line). The almost
parallel line corresponds to the bouncing solution for $u=0$, $\chi_0=1.155$.}
\end{figure} 

Inserting solution (\ref{u}) into
Eq.~(\ref{pt2}) in the limit $|T|\rightarrow\infty$, we obtain
\begin{equation}
P_T=\biggl(\frac{\chi_0}{T_0}+u\biggr)^2.
\label{pt13}
\end{equation}
Hence, the huge values of $P_T$ can be obtained from large values of $u$,
without any imposition on the parameters $\chi_0$ and $T_0$.

Let us calculate the constraints on the parameter
space in order to obtain a sensible model. As discussed in section II, one
should have $a_0^2/T_0\leq 1$ in order to have a reasonable probability
amplitude for $a_0$, and the curvature scale at the bounce should not
be very close  to the Planck length in order to avoid strong quantum gravitational effects.
I will also impose that $a_0>1$ in order to avoid transplanckian problems (see below), 
which implies that $a_0/T_0\leq 1$.

The curvature scale at the bounce reads
\begin{eqnarray}
\label{lscale}
L_{\rm bounce}&=& \frac{a_0 T_0 \sqrt{a_0[1+\cos(4ua_0)]}}{\sqrt{a_0[1+\cos(4ua_0)]+uT_0[4ua_0+\sin(4ua_0)]}}\nonumber\\
&\approx&\frac{a_0\sqrt{T_0}}{2u},
\end{eqnarray}
where in the last approximation I used that $uT_0/a_0>>1$, which follows from $u>>1$,
$a_0/T_0\leq 1$, and I assumed that $4ua_0\neq (2n+1)\pi$, in order to avoid 
$L_{\rm bounce}<<1$. Hence, as $a_0\leq T_0^{1/2}$, then 
\begin{equation}
L_{\rm bounce}\leq \frac{T_0}{2u}
\label{urlb}
\end{equation}
Demanding that $L_{\rm bounce}\geq 10^3$, than 
\begin{equation}
T_0 > u 10^3 >> 1.
\label{urt0}
\end{equation} 

One must check again whether one recovers the classical radiation
dominated evolution before nucleosynthesis. As before, I concentrate
on the case $w=1/3$, which implies that $\chi=a$ and $T=\eta$. I will
do this in two steps: first I check whether solution (\ref{u}) is valid
before nucleosynthesis, and then whether quantum effects are negligible there.

The approximation leading to solution (\ref{u}) requires that the argument
of the hyperbolic functions in Eq.~(\ref{exact}) be large, $\theta T/T_0>>1$,
which implies that $x=\eta/T_0>>(4u^2T_0-1)^{-1/2}\approx 1/(2uT_0^{1/2})$.
Hence, the values of the conformal time for which solution (\ref{u}) is reliable
are $\eta >> T_0^{1/2}/(2u)$.

Let us now verify whether solution (\ref{u}) is valid around the freeze-out
of neutrons, before nucleosynthesis. 
This will be true
if $\eta_{\rm f}>> T_0^{1/2}/(2u)$. However, 
$\eta_{\rm f}\approx 10^{22}/P_T^{1/4} = 
10^{22}/u^{1/2}$ which, when combined with $\eta_{\rm f}>> T_0^{1/2}/(2u)$, implies that
$10^{44}>>T_0/(4u)\geq L_{\rm bounce}$, where I used Eq. (\ref{urlb}). 
As the curvature scale around freeze-out of neutrons, $L_{\rm f}$, satisfies
$L_{\rm f}\approx 10^{44}$, this condition is just the reasonable
constraint that 
\begin{equation}
L_{\rm bounce}<<L_{\rm f}\approx 10^{44} .
\label{lbounce}
\end{equation} 
Hence, if the curvature scale
at the bounce is much smaller than the curvature scale around freeze-out of neutrons, than
solution (\ref{u}) must be valid at nucleosynthesis period.

Finally, one must check that in the regime where solution (\ref{u}) is valid 
we are already in the classical limit, even
though the above condition $x>>1/(2uT_0^{1/2})$ may still contain a
region were $x<<1$ because $u$ and $T_0$ are large. To prove this, note first 
that at $\eta>>T_0^{1/2}/(2u)$, the term $u\eta>>T_0^{1/2}/2$ dominates over $a_0\sqrt{x^2+1}$ in Eq.~(\ref{u}),
either for $x<<1$, because $a_0\leq T_0^{1/2}$, as for $x>>1$, because $a_0/T_0<<u$.
Hence, the quantum potential given by
\begin{equation}
Q:=- \frac{\partial^2 R}{4R\partial\chi^2}=:Q_1+Q_2^2,
\label{qp}
\end{equation} 
where 
\begin{widetext}
\begin{eqnarray}
\label{Q1}
Q_1&=&
-T_0\{[4T_0(\chi^2+u^2 T^2)-(T^2+T_0^2)]\cosh(\theta T/T_0)-8uT_0T\chi\sinh(\theta T/T_0)\nonumber\\
&+&[4T_0(\chi^2-u^2T_0^2)-(T^2-T_0^2)]\cos\theta + 8T_0^2\chi u\sin\theta\}
\{2(T^2+T_0^2)^2[\cosh(\theta T/T_0)+\cos\theta]\}^{-1},
\end{eqnarray}
and
\begin{equation}
Q_2=T_0\frac{X[\cosh(\theta T/T_0)+\cos\theta]-u[T\sinh(\theta T/T_0)-T_0\sin\theta]}
{(T^2+T_0^2)[\cosh(\theta T/T_0)+\cos\theta]},
\end{equation}
\end{widetext}
reads, around the Bohmian trajectory $a\approx u\eta$,
\begin{equation}
Q\approx \frac{1}{\cosh(2\theta T/T_0)}<<1,
\end{equation}
while the kinetic term of the
Hamilton-Jacobi-like equation (\ref{hamilton-jacobi}) is given by,
\be
\label{k}
\biggl(\frac{\partial S}{2\partial a}\biggr)^2 \approx u^2 >> 1.
\en

Note that near the bounce at $\eta\approx 0$, the kinetic term is almost null, while
the quantum potential is finite: quantum effects are dominant only very near the bounce.
The transition from quantum to classical regime should be around $\theta T/T_0\approx 1$,
where $Q\approx 1$ and solution (\ref{u}) is not reliable. A little bit later, when
$x=\eta/T_0 >(4u^2T_0-1)^{-1/2}\approx 1/(2uT_0^{1/2})$, and knowing from Eq. (\ref{urt0})
that $T_0 > 10^{61}$ because $u > 10^{58}$, we obtain that the scale factor at the beginning
of the classical regime is $a>10^{31}$ which is the minimum value required for the model
to reach the size of the observed Universe without needing any classical inflationary phase afterwards.
Note from figure 1 that the presence of the $u$ term in Eq. (\ref{u}) induces a much bigger
acceleration at the bounce in comparison with the solution with the same $a_0$ but without the $u$ term. 
It is this term which is the
responsible for the big value $a>10^{31}$ when the model enters the classical regime.

Concluding, solution (\ref{u}) reachs the standard cosmological model before 
nucleosynthesis, and can indeed describe the observed Universe in the radiation dominated phase
with parameters with reasonable relative probabilities.

In order to avoid the transplanckian problem for the scales of physical interest
today, $10^{54}<\lambda_{\rm today}^{\rm physical}<10^{60}$, one should have
$\lambda_{\rm bounce}^{\rm physical}$ corresponding to these scales not smaller than, 
say, $10^3$. As 
\be
\label{trans}
\lambda_{\rm bounce}^{\rm physical}=
\frac{a_0}{a_{\rm today}}\lambda_{\rm today}^{\rm physical},
\en
this problem can be avoided if
\be
\label{33}
\frac{a_{0}}{a_{\rm today}} > 10^{-51}  ,
\en
which implies that $a_0 > 10^9$.

Note that there is an upper limit for $a_{0}$ coming from
\be
L_{\rm bounce}\approx \frac{a_0 T_0^{1/2}}{2u}
\geq10^{64}\biggl(\frac{a_0}{a_{\rm today}}\biggr)^2 ,
\en
where I used that $a_0^2/T_0\leq 1$, $P_T \approx u^2 \approx a_{\rm today}^4 10^{-128}$.
The constraint $L_{\rm bounce}<<10^{44}$ (see Eq.(\ref{lbounce})) then
implies that
\be
\label{333}
\frac{a_0}{a_{\rm today}}<<10^{-10}.
\en
Hence, there is a large domain of values of $a_0$ where the transplanckian problem
can be avoided (see Eqs.(\ref{33},\ref{333})).

\section{Conclusion}

I have shown in this paper how a sufficiently big universe can emerge
from a quantum cosmological bounce, without needing any classical inflationary phase
afterwards to make it grow to its present size. This is caused by a huge
acceleration during the quantum bounce, which may be viewed as a quantum inflation.
These results were obtained from a moving gaussian function of the scale factor,
which is a solution of the Wheeler-DeWitt equation
coming from the canonical quantization of general relativity sourced
by relativistic particles. The solution is exact, there is no WKB approximation
involved here. Its value at $T=0$ yields reasonable relative probability amplitudes of having the
scale factor at the bounce with the value $a_0$ required to avoid any transplanckian
problem, and to allow that the curvature scale at the bounce be some few 
orders of magnitude greater than the Planck length, a region where one
can rely on this simple quantization scheme. In fact, as the maximum value
the curvature scale can have is at the bounce itself, one never reachs energy
scales where more involved quantum gravity theories, like string theory and
loop quantum gravity (see Ref.~\cite{nicolai} about issues concerning this
approach), must be invoked: the model is self-contained. 

There are two internal parameters of the wave functon which must be 
big in order to obtain a large classical universe from a quantum bounce.
The first one is the parameter $T_0$, the square root of the width of the
gaussian at the moment of the bounce (see Eq.~(\ref{initial2})), 
which must satisfy $T_0 > 10^{61}$ (in Planck unities). This value yields the sufficient large
value $a>10^{31}$ for the scale factor in the beginning of the classical
regime, and guarantees that the curvature scale at the bounce be some minimum
orders of magnitude greater than the Planck length in order for the Wheeler-DeWitt
equation I used be reliable. The other one is the velocity $u$ of the gaussian
along the scale factor axis which must satisfy $u > 10^{58}$ in order to yield
the amount of radiation we observe in the Universe today, without appealing
to some huge production of photons during the bounce.

From these considerations, one can see that the parameters emerging from
the quantum era of the Universe are not necessarily Planckian: they depend
also on the quantum state of the system, on the internal parameters of the
wave function of the Universe. Hence, it is not surprising that one may have
quantum gravity effects in large (when compared with the Planck length) 
Universes~\cite{big}, which could be dramatically seen in a big-rip~\cite{bigrip}.

However, one may ask why the internal parameters of the wave function we
obtained are so large. Note first that these are not coupling constants, but
parameters in the quantum state of the Universe. Hence, to answer this question, one should rely on some deep
understanding of quantum cosmology and/or new principles which are
not available today. Note, however, that the big value of $T_0$ leads to a widely spread
gaussian, and hence almost all scales at the bounce
are equally probable. This is a reasonable assumption about the wave function
of the Universe one can make: it should not intrinsically select any preferable 
scale at the bounce without any special reason. 
Concerning the $u$ variable, its large value implies that the peak of the initial 
wave packet moves very fast towards large scale factors, which induces a 
large universe. Perhaps some version of the Anthropic Principle could justify
the preference for large classical universes, and as consequence for a large $u$,
but I think the important message here is the possibility of 
obtaining a large universe from a huge acceleration of the scale factor in the far past, 
whose origin differs fundamentally from those considered in usual inflationary scenarios. 

In future publications, we will calculate the evolution of linear
quantum perturbations and particle production on these quantum 
backgrounds, as in Ref.~\cite{PPN4}, and compare the results with observations.
 
As a final remark, I would like to repeat a comment we
made elsewhere~\cite{noinf}: in contradistinction
with models in which time begins, there is no point on asking what is
the probability of appearance of some particular eternal model out of
nothing. Contrary to usual perspectives, one can as well assume
existence to be conceptually prior to non-existence, \ie existence
itself may not be deserving explanation.  This is the idea underlying
our category of models: the Universe always existed and its
``appearance'' is thus a non question.

\section*{Acknowledgements} I would like to thank CNPq of Brazil for
financial support. I very gratefully acknowledge various enlightening 
conversations with Felipe Tovar Falciano,
Patrick Peter, Emanuel Pinho, 
and specially Andrei Linde, whose critcisms inspired
this work. I also
would like to thank CAPES (Brazil) and COFECUB (France) for partial
financial support.

\end{document}